\documentclass[%
 aip,
rsi,%
 amsmath,amssymb,
 reprint,%
]{revtex4-1}

\usepackage{graphicx}% Include figure files
\usepackage{graphics}
\usepackage{dcolumn}% Align table columns on decimal point
\usepackage{bm}% bold math
\usepackage{color}
\begin{document}

\preprint{AIP/123-QED}

\title{Slow Relaxations of Chemically Confined Hydration Layers near
Lipid Bilayers: Dynamical Heterogeneities above Supercooling}%:\\with Forced Linebreak\footnote{Error!}}% Force line breaks with \\
% \thanks{Footnote to title of article.}

% \author{A. Author}
%  \altaffiliation[Also at ]{Physics Department, XYZ University.}%Lines break automatically or can be forced with \\
\author{Abhinav Srivastava} %
 \affiliation{Department of Chemistry, Indian Institute of Technology Jodhpur, Jodhpur 342037, India.}
%\\This line break forced with \textbackslash\textbackslash

 \author{Smarajit Karmakar}%$\ddagger$}
 \affiliation{Center for Interdisciplinary Sciences, Tata Institute of Fundamental Research, Hydreabad 500107, India.}
 
  \author{Ananya Debnath}%$\dagger$}
  \email{ananya@iitj.ac.in}
  \affiliation{Department of Chemistry, Indian Institute of Technology Jodhpur, Jodhpur 342037, India.}

\date{\today}

\begin{abstract}

A hydrated 1,2-dimyristoyl-sn-glycero-3-phosphorylcholine (DMPC) lipid membrane is investigated 
 using an all atom molecular dynamics simulation at 308K to find out the physical sources of 
 universal slow relaxation of hydration layers. Continuously residing interface water (IW)
 hydrogen bonded to each other and concertedly to different moieties of lipid heads are identified. 
 The non-gaussian parameter of all IW show a crossover from cage vibration to translational diffusion.
 A significant non-gaussianity is observed for the IW prevailing large length correlations 
 in translational van Hove functions. Two time-scales for the ballisitic motions and hopping 
 transitions are obtained from the self intermediate scattering functions of the IW with 
 an additional long relaxation which disappears for the BW. This is attributed to the coupled 
 dynamics of IW cages hydrogen bonded to lipid heads. Our calculations reveal that the water 
 near membranes are slowed down due to dynamical heterogeneities above room temperature and 
 have implications to bioprotection mechanism at freezing conditions.

%
%Valid PACS numbers may be entered using the \verb+\pacs{#1}+ command.
\end{abstract}

\maketitle

%\keywords{Hydrogen bonding, reorientation relaxation, bound waters}

\section{Introduction}

Water is the most abundant and the principle constituent of any living organism \cite{Eisenberg}.
In biological systems, water is confined between biomolecular assemblies where water dynamics is
relevant in membrane functioning and cytoskeletal organizations \cite{Pavel_JPCL_15,Lingwood_Science_2010,Munro_cells_03}.
Recently, protein displacements in live cell membranes are found to have robust exponential tails
due to the underlying dynamical heterogeneities \cite{He_nature_2016,MadanRao_Cell_2012}.
The nano-scale heterogeneity in membrane dynamics is believed to play the dominant role in various cellular
processes such as signal transduction,
matter transport and enzymatic activities in cells \cite{Mouritsen_BioEssays_1992}.
Dynamical heterogeneities in fluid membranes are observed
on the scale of 80-150 nm with super-resolution stimulated emission depletion microscopic and flourescence correlation 
spectroscopic techniques, even in absence of cholesterols \cite{Jaydeep_Soft_Matter_2017}.\\
Although, dynamical heterogeneity in membranes is not well-studied yet, the obvious question comes if
dynamical heterogeneity in the membrane is coupled with the dynamics of its hydration layer \cite{Ananya_PRL_2013,Ananya_JCP_2018}.
In the last decade, with a major advancement in computer simulation techniques, water near soft interfaces
are found to have very slow relaxation time \cite{Ji_Science01,Mukherjee_JPCB_15,Sundaram_PRL_02,
Bagchi_JCP_13,Zewail_JPCB_2002}. 
However, the origin and mechanism of slow relaxation time have not been 
analyzed in a systematic manner with microscopic details till date. If 
universal slow hydration dynamics is due to the breaking of tetrahedral hydrogen bond network near interfaces, is
their dynamics dependent on the chemical nature of the confinement present in the interfaces? 
 Confined diffusion of nano-particles are found to 
follow non-gaussian statistics even at the long time Brownian stage \cite{Xue_JPCL_16}. Dynamical heterogeneity time and length scales
are analyzed for the ionic liquids and water binary mixtures where jump motions are evident from the non-gaussian distributions \cite{Ranjit_JPCB_2015,Ranjit_JCP_2016}.
Activated hopping is known to facilitate the random diffusion in supercooled liquids \cite{Sarika_PNAS_08}. Water confined near protein surface
are shown to exhibit a dynamic crossover with a breakdown of Stokes Einstein relation at supercooled temperatures \cite{Chen_JPCB_2010}.
Water confined near silica hydrophilic pores at room temperature reveal two different dynamical
regimes similar to supercooled bulk water \cite{Gallo_JCP_2000,Gallo_PRL_2000}.
The open question remains, do confined water at room temperature dynamically behave very similar to 
supercooled bulk water? If yes, what are the sources of such distinct dynamics  \cite{Funel_JML_1998,Bilge_JACS_2011,Chen_PRE_1996,SH_Chen_PRL_1996,Eugene_PRL_2006,Biman_JPCB_09}?\\
In this article we present evidences of dynamical heterogeneities in chemically confined
water near DMPC lipid headgroups at temperature 
 ($308$ K) well above supercooling. The influence of chemical nature on the hydrogen bond dynamics of interface water
 is systematically analyzed to find out the origin of their slow relaxation rates. Our calculations provide the microscopic mechanism responsible
 for such behavior at room temperature with their implications and potential applications.\\

\section{Simulation Details}
An all atom molecular dynamics simulation is carried out for 128 DMPC molecules in presence of 5743 TIP4P/2005 \cite{JLF_JCP_05} water molecules
using previously equilibrated DMPC system at 308 K\cite{Ananya_JCP_2018}. 
Force field parameters for DMPC are obtained using Berger united 
atom force field \cite{Berger_Biophys_97,Cordomí_JCTC_12}. An NPT run is carried out for 100 ns with a 2 fs time step. 
The system is equilibrated at 308 K using velocity rescaling method 
with a coupling constant of 0.5 ps. The pressure is maintained at 1 bar using semi-isotropic pressure coupling by 
Berendsen pressure coupling\cite{Berendsen_JCP_84} 
with a coupling constant of 0.1 ps. Coulombic and van der Waal interactions were cutoff at 1 nm. 
Long range interactions are corrected using particle mesh Ewald\cite{Ulrich_JCP_95, Allen, Darden_JCP_93} 
method with a 4 nm grid size. Periodic boundary conditions are applied in all three directions. 
%Trajectories were collected after every 200ps.\\
An NVT simulation is performed for 1.9 ns with a 0.4 fs time step where the last 1 ns run is analyzed for water dynamics. 
Parameters for temperature coupling, cutoff distances and long range interactions are kept same as in the previous run \cite{Ananya_JCP_2018}. 
Trajectories are collected at every 10 fs. The simulation box-length for the hydrated DMPC lipid is 
6.24 nm along x and y directions and 7.95 nm along z direction.
To compare the dynamics of interfacial water with bulk water (BW), a NVT run is carried out for a box of 851 TIP4P/2005 water molecules
using the same parameters as in DMPC-water system.
for 2 ns with a 2 fs time step in an NPT ensemble with the same set of parameters as in the hydrated DMPC. Next a NVT run is carried out 
for 200 ps with a 0.4 fs time step. The 100 ps NVT run is further extended till 200 ps where we have used last 100 ps data for analysis.
The box length for BW is 3.69 nm along x and y and 1.84 nm along z directions. 
Trajectories are collected at every 10 fs. 
All simulations are carried out using Gromacs 4.6.5\cite{RENARDUS_PC_93,Hess_MMA_01,Erik_JCTC_08,JCH_JCC_05,Hess_2013,Drunen_CPC_95}.

Figure \ref{snapshot} (a) and (b) show the bilayer and a single DMPC
molecule present in the bilayer respectively. To decouple the contribution of bulk water (BW) to the dynamical properties
of hydration layers, interface water are classified based on geometric definitions. If a water molecule continuously resides in a layer which is $\pm3\AA $ away from the location of 
the head group density of DMPC, the molecule is identified as interface water (IW) \cite{Ananya_PRL_2013,Ananya_JCP_2010}. The inset in figure \ref{RDF-lifetime}
shows the number of IW with respect to their lifetimes. IW has a bi-exponential lifetime dependence which is indicative of the presence of two charasteristics relaxation 
time-scales. IW which stay continuously in the hydration layer for 100 ps are found to provide reasonable statistics to calculate dynamical quantities and used for further analyses.
If one IW of 100 ps lifetime is hydrogen bonded to another IW with same lifetime, it is referred to as IW-IW.
Additionally, if one pair of IW-IW is concertedly hydrogen bonded to the carbonyl (CO), phosphate (PO) or glycerol (Glyc) moeity of lipid heads (shown in
different colors in figure \ref{snapshot} b)), the
IW-IW is referred to as IW-CO, IW-PO or IW-Glyc respectively \cite{Ananya_JCP_2018}. The approach clearly de-constructs the contributions of different chemical environments on the dynamical behaviors of interface water 
which is lacking if water molecules are characterized depending on their vertical positions along the bilayer normal. To obtain the dynamics of the slowest IW with reasonable
statistics, another class of IW is chosen which stay continuously in the hydration layer for 400 ps and is referred to as IW$_{CR400}$. 

\section{Results and Discussions}
\begin{figure}
 \includegraphics[width=\linewidth]{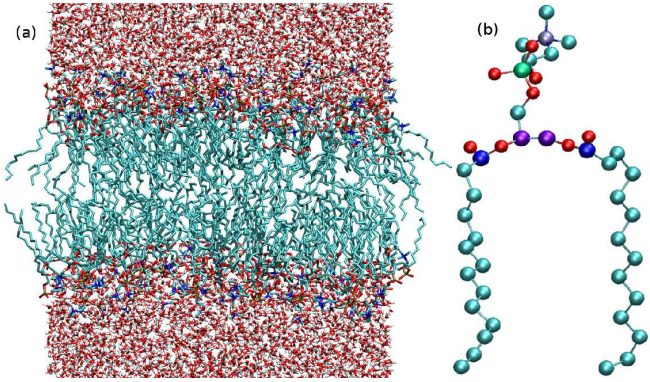}
\caption{a) Snapshot of a DMPC bilayer in presence of interface water, b) snapshot of a single DMPC molecule showing different
moieties in different colors, blue: carbonyl carbon, red: oxygen, green: phosphorus of phosphate, violet: glycerol
carbon. }
 \label{snapshot}
\end{figure}

\begin{figure}
 \includegraphics[width=\linewidth]{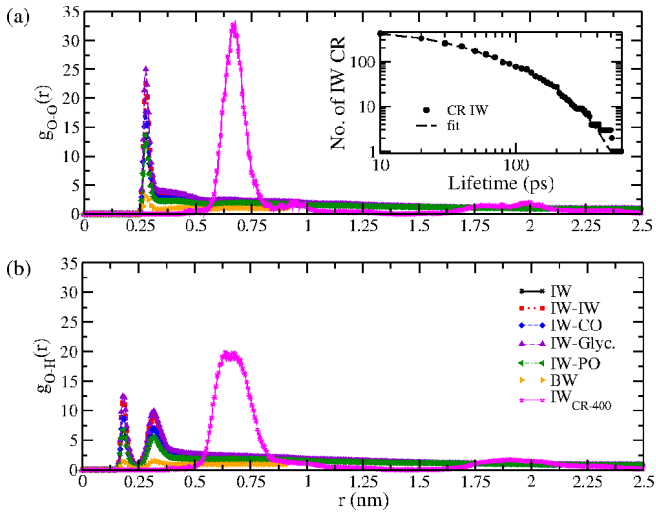}
\caption{RDF for different classes of IW and BW between a) oxygen-oxygen and b) oxygen-hydrogen atoms. 
Inset: Lifetime of interfacial waters continuously residing (CR) within $\pm 3 \AA$ away from the peak of the density profile
of nitrogen atoms of lipid heads.}
 \label{RDF-lifetime}
\end{figure}

\begin{figure}
 \includegraphics[width=\linewidth]{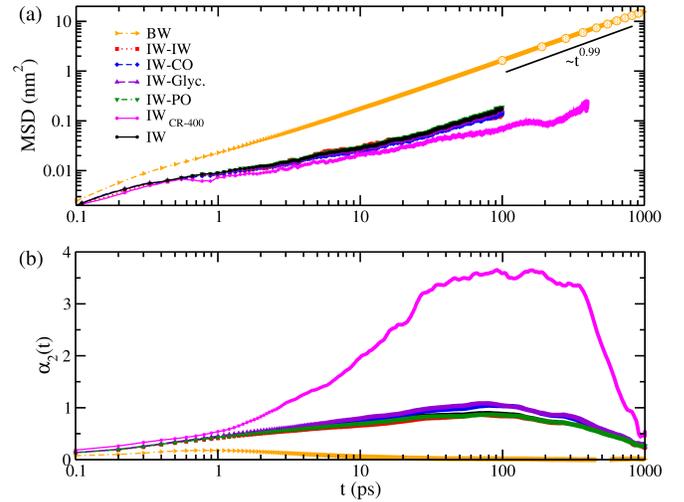}
 \caption{a) Translational mean square displacement, b) NGP for all classes of IW and BW
 show a crossover from $\beta$-relaxation to $\alpha$-relaxation at the same time-scale when respective sub-diffusive regimes
 are left for diffusion.}
 \label{MSD-NGP}
\end{figure}

The radial distribution functions ($g(r)$) between two oxygen atoms and oxygen and hydrogen atoms of different classes of IW are shown in figure \ref{RDF-lifetime}
(a) and (b) respectively. The highest amplitude of IW$_{CR400}$ with respect to remaining classes of water indicates that the IW$_{CR400}$ have the lowest potential energies 
due to the breakdown of tetrahedral networks while forming hydrogen bonds to the lipid head moieties .
To analyze the dynamical nature of bound waters, translational mean square displacement (MSD, $\frac{1}{N}\sum_{i=1}^N<({\bf r_i}(t)-{\bf r_i}(0))^2>$) is calculated for all classes
of water. Figure \ref{MSD-NGP} (a) shows that the BW follows diffusive behavior at longer time where all classes of IW
remain subdiffusive due to the trapping in a cage formed by the neighboring hydrogen bond networks. IW$_{CR400}$ exhibit the slowest MSD due to the longest confinement lifetime.
To understand the origin of subdiffusive nature of the IW, non-Gaussian paramater (NGP, $\alpha_2$) are calculated 
using the equation, $ \alpha_{2}(t) = \frac {3<\Delta {\bf r} (t)^{4}>} {5<\Delta {\bf r} ( t)^{2}>^{2}} -1$ \cite{Stillinger_JPCM_2005,Hansen_book}
(figure \ref{MSD-NGP}). The NGP for the BW 
reaches a maximum within a small time, 0.9 ps and then decays asymptotically to zero confirming the gaussian diffusion. 
The values of NGP for different classes of IW, start increasing
to much higher amplitudes than that for the BW and reach to maxima in between $\beta$ and $\alpha$-relaxation time scales. 
The slow decays after the peak of NGP are due to the release of the IW from the 
respective cages via diffusion. Similar cross-over from cage to translational diffusive regime at the
peak of NGP are found for Brownian particles in a periodic effective field \cite{MAJ_PRE_2007}. Since different
classes of IW which continuously reside for 100 ps in the hydration layer, leave the layer after their confinement lifetimes,
the maxima of their respective NGP are nearly at 100 ps (see Table 1)
% in supplementary information). 
IW-Glyc being burried deepest in the hydrophobic core of lipid 
(figure \ref{snapshot} b)) show the highest $\beta$-relaxation among IW which are continuously residing for 100 ps. Interestingly, IW$_{CR400}$ exhibit very strong non-gaussian behavior 
for longer period of time compared to other classes of water molecules and decay after ~340 ps which is again close to their
confinement lifetimes. A very slow $\alpha$-relaxation of IW$_{CR400}$ is indicative of their
slow diffusion towards far-interface region. The broad $\beta$-relaxations of all classes of IW signify a strong structural arrest by their
surrounding molecules similar to supercooled liquids exhibiting dynamical heterogeneity \cite{Bilge_JACS_2011,Walter_Kob_PRL_1997}. The transition
from the $\beta$-relaxation to the $\alpha$-relaxation occurs at similar spatio-temporal scale where the respective MSD leave the sub-diffusive regime.\\

\begin{figure}
 \includegraphics[width=\linewidth]{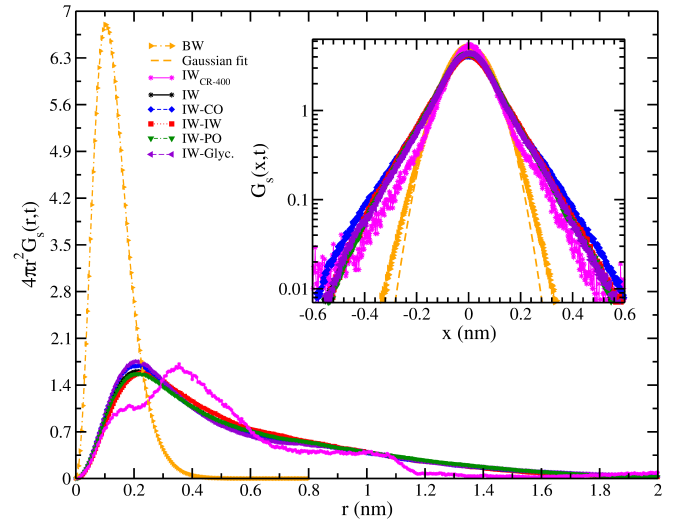}
 \caption{Translational self part of van Hove correlation function for all classes of IW and BW. All IW pertain large length correlation than the BW.
 Inset: van Hove correlation function along x direction for all classes of IW and BW. IW show a stong deviation from gaussianity.}
 \label{vanhove}
\end{figure}

\begin{figure}
 \includegraphics[width=\linewidth]{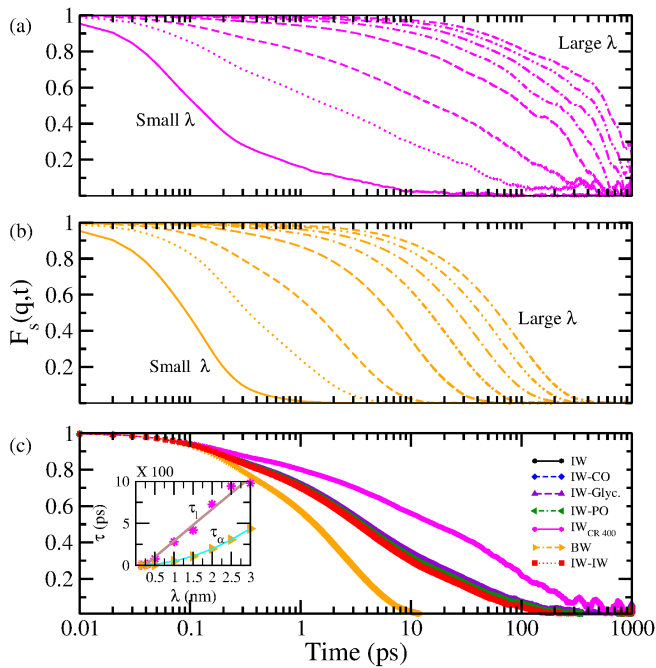}
 \caption{ Self intermediate scattering function, $F_s({\bf q},t)$ for a) IW$_{CR400}$ and b) BW at different $ \lambda$  
 ranging from $\lambda =0.15$ nm to $\lambda =3.00$ nm ( ${\bf q} = \frac{2\pi}{ \lambda}$). 
 Inset: $\alpha$-relaxation dependance on $ \lambda$. Symbols: $\tau_\alpha$ and $\tau_l$ for IW$_{CR400}$ and BW respectively; lines : fitting.}
 \label{fskt}
\end{figure}

\begin{table}[ht] 
\caption{Relaxation time-scales of different IW and BW as obtained from NGP. For BW, $\alpha_{2}$ decays to $0$
following Fickian dynamics where remaining IW show strong non-gaussian behavior.}
\centering
\begin{tabular}{lcc}
\\ \hline\hline
Region              & $ t$ \\
                    &     (ps)   \\        
\hline
IW                  &  78.70      \\ 
IW-IW               &  70.10      \\
IW-CO               &  80.10      \\
IW-Glyc            &  82.00      \\
IW-PO               &  78.40      \\
IW$_{CR400}$       & 344.80      \\
BW                  &   0.90      \\
\hline
\label{NGP_t}
\end{tabular}
\end{table}
For gaining deeper insights in the dynamical evolution of IW associated with lipid moieties, self part of radial van Hove 
correlation function \cite{Schmidt_JCP_2010} is calculated via, $G_{s}({\bf r},t)= \frac{1}{N} \bigg\langle\sum_{i=1}^{N} \delta({\bf r}+{\bf r}_{i}(0)-{\bf r}_{i}(t)\bigg\rangle$. 
Figure \ref{vanhove} shows the radial van Hove correlation function with a time interval
corresponding to the $\beta$-relaxation of NGP where the dynamical heterogeneity is most significant for all classes of water. All
classes of IW show very strong deviations from guassianity with stretched exponential decays where the BW follow gaussian behavior. 
Due to the chemical confinement of different classes of IW
in the vicinity of lipid heads, there is a strong correlation pertaining
to longer length-scales which decays at much smaller length scale for the BW. Among different classes of IW which continuously
reside in the hydration layer for 100 ps, the van Hove correlation function of IW-Glyc have the maximum amplitude consistent with the
nature of the respective NGP. Importantly, the van Hove correlation function of IW$_{CR400}$ has two peaks and a shoulder since IW$_{CR400}$ might follow a 
rattling or hopping mechanism in the specific
cages of another neighboring IW for longer lifetimes.
However, their correlations decay at a smaller length scale than that for the other classes of IW.
The inset in the figure \ref{vanhove} represents one dimensional van Hove correlation function where the BW
follow gaussian dynamics and all classes of IW show larger deviations from gaussianity
via exponential tails. Similar exponential tails have been manifested as an established behavior of dynamical heterogeneity in supercooled
glass forming liquids \cite{Pinaki_PRL_07,Siladitya_JCP_14}.\\
% The origin of exponential tails for the IW is the waiting time distributions for
% the subsequent jumps between cages formed by the neighboring hydrogen bond networks. The exponential tails of van Hove functions at different 
% temperatures can provide information on breakdown of Stokes-Einstein relation with
% particles of different mobilities \cite{Siladitya_JCP_14}.\\
Since self intermediate scattering function (SISF, $F_s({\bf q},t)$) is another universal feature of dynamical heterogeneity, two dimensional $F_s({\bf q},t)$ is calculated by,
 $F_{s}({\bf q},t) = \frac{1}{N} \bigg \langle \sum_{i=1}^{N} Cos({{\bf q}.[{\bf r}_i(t) - {\bf r}_i(0)] }) \bigg \rangle.$
${\bf q}$ is obtained from the location of the first peak of the $g(r)$ via ${\bf q} = \frac{2\pi}{ \lambda}$ where $ \lambda$ is the wave length. 
To obtain the wave-vector dependence on the $\alpha$-relaxation times, SISF are calculated at different
values of $ \lambda$. 
Figure \ref{fskt} (a) and (b) show the behavior of SISF at different $ \lambda$ for the IW$_{CR400}$ and the BW. 
Although the $\beta$
and $\alpha$-relaxation times of the IW and the BW are not prominently disparate by characteristics boson peaks as observed for the supercooled liquids \cite{Hansen_book},
SISF very clearly scales up in different time regimes. The  
wave length dependence of $\alpha$-relaxation time-scales ($\tau_\alpha$) is generally characterized by the exponential decay of 
$F_{s}({\bf q},t)=exp(-D{\bf q}^2t)$ at all wave lengths at room temperature. If diffusion coefficients ($D$) and relaxation
times of IW follow a distribution at room temperature due to the dynamical heterogeneity, the quadratic dependence of $ \lambda$ to the long relaxation time will 
not be followed anymore. To check that, the long $\tau$ are extracted from the data
in figure \ref{fskt} (a) and (b) for the IW$_{CR400}$ and the BW
respectively. The inset in figure \ref{fskt} (c) shows much slower $\alpha$-relaxation time-scales for the IW$_{CR400}$ in comparison to the BW (the data for the BW is multiplied by a factor of $5$
for a better comparison). Interestingly, the $\tau_\alpha$ of the BW follow quadratic
dependence to the $ \lambda$ which changes to a linear dependence for the IW$_{CR400}$.
This is probably because the movement of confined water are heterogeneous in nature at all length-scales due to the breakdown of Stokes-Einstein relation at room temperature
\cite{Srikanth_PRL_17}.
$\lambda=0.5$ nm is chosen to understand the nature of characteristics relaxation time-scales of all classes of IW and BW (figure \ref{fskt} (c)).
All IW which continuously reside at the hydration layer for 100 ps show much slower relaxation compare to the BW.
Our BW data are fitted best to two relaxation time-scales using, 
$F_s({\bf q},t) = (1-f_Q) \exp [-(\frac{t}{\tau_s})^2] + f_{Q}\exp [-(\frac{t}{\tau_{\alpha}})^{\beta_{\alpha}}]$.
$f_Q$ is known as the Debye-Waller factor, $\tau_s$ is the time-scale for ballistic motion and $\tau_{\alpha}$ is 
the relaxation time-scale of the cage \cite{Gallo_JPCL_2011}. The stretched exponential in the equation is known as Kohlrausch-William-Watt (KWW) function \cite{Gallo_JCP_2016}
 where $\beta_{\alpha}$ can be correlated with the breakdown of Stokes-Einstein relation for supercooled liquids \cite{Smarajit_JStatMech_16}.
However, all classes of IW are not fitted to the previous equation due to
the appearance of a long time tail very similar to that for the IW near proteins \cite{Gallo_JCP_2016}. The SISF of IW can be accounted with three relaxation time-scales where one more stretched parameter is added to the KWW function as 
$F_{s}({\bf q},t) = (1-f_Q-f'_Q) \exp [-(\frac{t}{\tau_s})^2]  + 
f_Q\exp[-(\frac{t}{\tau_{\alpha}})^{\beta_{\alpha}}] 
+ f'_Q\exp [-(\frac{t}{\tau_l})^{\beta_l}]$.
$\tau_l$ and $\beta_l$ are longer relaxation time and stretching parameter respectively. Although such stretched long relaxation
is not so common in glass-former liquids, $\tau_l$ for the IW near protein is known to follow Arrhenius dependence on temperatures with a cross-over
due to the protein structure fluctuations \cite{Gallo_JCP_2016}.
Table 2 
% in supplementary information 
shows the relaxation time-scales and stretching parameters for all IW and the BW near the DMPC.
$\tau_s$ and $\tau_{\alpha}$ of all classes of IW which continuously reside in the hydration layer for 100 ps, show 
similar time-scales as that of the BW and consistent with the time-scales obtained
for hydration water near protein or sugar \cite{Gallo_JCP_2016,Gallo_JPCL_2011}. The IW-CO/IW-Glyc exhibit slowest $\tau_s$ and $\tau_{\alpha}$ 
since they are buried deepest in the hydrophobic region of lipids. Interestingly, the $\tau_s$ and the $\tau_{\alpha}$ for IW$_{CR400}$
are 10 orders of magnitude slower than the remaining classes of water. However, all classes of water follow a long time
stretched exponential tail (described by $\tau_l$) which is clearly absent in the BW. Notably, the $\tau_l$ for IW$_{CR400}$
is 4-6 times larger than that for the remaining classes of IW. The emergence of $\tau_l$ is attributed to the very slow
relaxation of the IW arrested in a cage of hydrogen bond networks formed near the lipid head groups. Since the time-scale of $\tau_l$
for the IW matches with the reorientation and translation time-scale of lipid heads (which is ~10-100 ns) \cite{Das_JCP_13},
a coupling between lipid head dynamics to the IW hydrogen bond dynamics can be the source of such long relaxations.\\

 \begin{table}[ht] \label{sisf-parameters}
\caption{Fitting parameters of the SISFs for all classes of interfacial and bulk water. Correlation coefficients were \textgreater 0.99.}
\centering
\begin{tabular}{lcccccccc}
\\ \hline\hline
Region                & $\tau_s$ & $f_Q$ & $\tau_{\alpha}$ & $\beta_{\alpha}$ & $f'_Q$ & $\tau_l$ & $\beta_l$\\
                      &     (ps)       &         &    (ps)         &                  &             &   (ps)        &               \\
\hline
IW                    & 0.29           &  0.23   &    2.87         &      0.99        &    0.71     &   15.26       &      0.48     \\
IW-IW                 & 0.21           &  0.36   &    2.69         &      0.92        &    0.55     &   20.93       &      0.59     \\
IW-CO                 & 0.32           &  0.21   &    2.86         &      0.99        &    0.75     &   15.03       &      0.46     \\
IW-Glyc              & 0.31           &  0.20   &    2.74         &      0.92        &    0.77     &   15.12       &      0.45     \\
IW-PO                 & 0.29           &  0.26   &    2.69         &      0.91        &    0.69     &   15.15       &      0.48     \\
IW$_{CR400}$      & 3.72           &  0.63   &    26.29        &      0.33        &    0.23     &   83.59       &      0.89     \\
BW                    & 0.24           &  0.88   &    2.53         &      0.93                                                      \\
\hline
\end{tabular}
\end{table}

\section{Conclusions}
In summary, the article provides evidences of spatio-temporal heterogeneities in interface water near lipid membranes
at temperature well above the glass transition temperature using all-atom molecular dynamics simulations.
Since IW$_{CR400}$ has the longest confinement lifetime, more profound dynamical heterogeneities
are observed for these molecules. Although dynamical heterogeneities persist for all classes of IW near the lipid membranes, the magnitude
of the heterogeneities are largely dictated by the hydrogen bond partner of lipid head moities.
IW-PO, IW-CO, IW-Glyc are interface water molecules which are hydrogen bonded among each other and concertedly hydrogen bonded to
PO, CO, Glyc of lipid heads. So for these cases, the dynamics are essentially probed for the hydrogen bonds between two 
IW molecules, although the hydrogen bond partners are different for IW-CO to IW-PO to IW-Glyc. Thus, the chemical nature of the PO, CO or Glyc of lipid heads
have less influence on the IW-IW hydrogen bond dynamics. On the other hand, IW$_{CR400}$ include only those interface water
which have formed hydrogen bonds directly to the lipid head moieties: PO, CO, Glyc and remain
intact for the entire 400 ps confinement lifetime. Thus the lipid partners have more influences on the dynamics
of the respective IW$_{CR400}$. Since these IW have formed hydrogen bonds to PO or CO or Glyc simultaneously, their interactions
are more heterogeneous in nature. The heterogeneous interactions within IW$_{CR400}$ may lower down the potential
energy of the respective cages compared to the ones having the homogeneous interactions very similar to the supercooled Lennard-Jones liquids \cite{Sharon_PRL_1997}. 
The heterogeneous interactions of IW$_{CR400}$ can generate a distribution of relaxation times which might be slower than the IW of remaining kind. 
Thus our calculations reveal for the first time that the slow relaxations of chemically confined water molecules near lipid membranes are originated from
dynamical heterogeneities at a temperature well above supercooling.\\
Additionally, our analysis shows strong signatures of coupling of lipid dynamics to hydration layer dynamics
contributing to the dynamical heterogeneities which merits further rigorous investigations. This will have strong implications on 
the dynamics of lipid rafts, skeleton fences \cite{Lingwood_Science_2010,Munro_cells_03,Ritchie_Biophy_2005} to understand if the membrane organizations
are driven by equilibrium processes \cite{He_nature_2016}. Moreover, it will be interesting to find out the length-scale dependence of
dynamical heterogeneity using the block analysis of van Hove function and four point susceptibility. Importantly, if the information of length-scale of dynamic heterogeneities is embedded to the 
length-scale of the lipid phase transition, the analysis will allow in predicting a length-scale for the phase transition.
Since confined biological water near membranes at room temperature exhibit similar dynamics as the 
super-cooled bulk water, this can enhance our understanding on the mechanisms of bioprotection during freezing stresses and have similar applications as
on cryo-preservations, but at room temperature.
At the same time, our findings raise few more questions: 
% how strong the lipid dynamics coupled to the water dynamics? Is it relevant to
% the phase transitions of lipid bilayers? 
Is the coupling between lipid and hydration
dynamics kinetic or thermodynamic in nature? Do interface water with large confinement lifetime have intermolecular energy flow via mode-coupling to keep the 
bonds intact? Is the nature of their motions cooperative? \\
\section{Acknowledgement}
AD is thankful to the project IITJ/SEED/20140016 for financial support. \\

\providecommand{\refin}[1]{\\ \textbf{Referenced in:} #1}

% \newpage

%\bibliographystyle{unsrt}
% \bibliography{paper_arxiv}

\begin{thebibliography}{0}%
\makeatletter
\providecommand \@ifxundefined [1]{%
 \@ifx{#1\undefined}
}%
\providecommand \@ifnum [1]{%
 \ifnum #1\expandafter \@firstoftwo
 \else \expandafter \@secondoftwo
 \fi
}%
\providecommand \@ifx [1]{%
 \ifx #1\expandafter \@firstoftwo
 \else \expandafter \@secondoftwo
 \fi
}%
\providecommand \natexlab [1]{#1}%
\providecommand \enquote  [1]{``#1''}%
\providecommand \bibnamefont  [1]{#1}%
\providecommand \bibfnamefont [1]{#1}%
\providecommand \citenamefont [1]{#1}%
\providecommand \href@noop [0]{\@secondoftwo}%
\providecommand \href [0]{\begingroup \@sanitize@url \@href}%
\providecommand \@href[1]{\@@startlink{#1}\@@href}%
\providecommand \@@href[1]{\endgroup#1\@@endlink}%
\providecommand \@sanitize@url [0]{\catcode `\\12\catcode `\$12\catcode
  `\&12\catcode `\#12\catcode `\^12\catcode `\_12\catcode `\%12\relax}%
\providecommand \@@startlink[1]{}%
\providecommand \@@endlink[0]{}%
\providecommand \url  [0]{\begingroup\@sanitize@url \@url }%
\providecommand \@url [1]{\endgroup\@href {#1}{\urlprefix }}%
\providecommand \urlprefix  [0]{URL }%
\providecommand \Eprint [0]{\href }%
\providecommand \doibase [0]{http://dx.doi.org/}%
\providecommand \selectlanguage [0]{\@gobble}%
\providecommand \bibinfo  [0]{\@secondoftwo}%
\providecommand \bibfield  [0]{\@secondoftwo}%
\providecommand \translation [1]{[#1]}%
\providecommand \BibitemOpen [0]{}%
\providecommand \bibitemStop [0]{}%
\providecommand \bibitemNoStop [0]{.\EOS\space}%
\providecommand \EOS [0]{\spacefactor3000\relax}%
\providecommand \BibitemShut  [1]{\csname bibitem#1\endcsname}%
\let\auto@bib@innerbib\@empty
%</preamble>
\end{thebibliography}%


\begin{thebibliography}{10}

\bibitem{Eisenberg}
Eisenber,~D.;\ \ Kauzmann,~W. \textit{The Structure and Properties of Water;}
  Oxford University Press: New York: 1969.

\bibitem{Pavel_JPCL_15}
Jungwirth,~P. \textit{The Journal of Physical Chemistry Letters} \textbf{2015,}
  \textsl{6,} 2449--2451.

\bibitem{Lingwood_Science_2010}
Lingwood,~D.;\ \ Simons,~K. \textit{Science} \textbf{2010,} \textsl{327,}
  46--50.

\bibitem{Munro_cells_03}
Munro,~S. \textit{Cell} \textbf{2003,} \textsl{115,} 377--388.

\bibitem{He_nature_2016}
He,~W.;\ \ Song,~H.;\ \ Su,~Y.;\ \ Geng,;\ \ Ackerson,~B.~J.;\ \ Peng,~H.~B.;\
  \ Tong,~P. \textit{Nature Communications} \textbf{2016,} \textsl{7,} 11701.

\bibitem{MadanRao_Cell_2012}
Gowrishankar,~K.;\ \ Ghosh,~S.;\ \ Saha,~S.;\ \ Rumamol,~C.;\ \ Mayor,~S.;\ \
  Rao,~M. \textit{Cell} \textbf{2012,} \textsl{149,} 1353 - 1367.

\bibitem{Mouritsen_BioEssays_1992}
Mouritsen,~O.~G.;\ \ J{\o}rgensen,~K. \textit{BioEssays} \textbf{1992,}
  \textsl{14,} 129--136.

\bibitem{Jaydeep_Soft_Matter_2017}
Roobala,~C.;\ \ K.,~J.~B. \textit{Soft Matter} \textbf{2017,} \textsl{13,}
  4598--4606.

\bibitem{Ananya_PRL_2013}
Debnath,~A.;\ \ Ayappa,~K.~G.;\ \ Maiti,~P.~K. \textit{Physical Review Letters}
  \textbf{2013,} \textsl{110,} 018303-1--5.

\bibitem{Ananya_JCP_2018}
Srivastava,~A.;\ \ Debnath,~A. \textit{The Journal of Chemical Physics}
  \textbf{2018,} \textsl{148,} 094901.

\bibitem{Ji_Science01}
Ji,~M.;\ \ Odelius,~M.;\ \ Gaffney,~K.~J. \textit{Science} \textbf{2010,}
  \textsl{328,} 1003--1005.

\bibitem{Mukherjee_JPCB_15}
Das,~S.;\ \ Biswas,~R.;\ \ Mukherjee,~B. \textit{The Journal of Physical
  Chemistry B} \textbf{2015,} \textsl{119,} 274--283.

\bibitem{Sundaram_PRL_02}
Balasubramanian,~S.;\ \ Pal,~S.;\ \ Bagchi,~B. \textit{Physical Review Letters}
  \textbf{2002,} \textsl{89,} 115505-1--4.

\bibitem{Bagchi_JCP_13}
Biswas,~R.;\ \ Furtado,~J.;\ \ Bagchi,~B. \textit{The Journal of Chemical
  Physics} \textbf{2013,} \textsl{139,} 144906-1--11.

\bibitem{Zewail_JPCB_2002}
Pal,~S.~K.;\ \ Peon,~J.;\ \ Bagchi,~B.;\ \ Zewail,~A.~H. \textit{The Journal of
  Physical Chemistry B} \textbf{2002,} \textsl{106,} 12376-12395.

\bibitem{Xue_JPCL_16}
Xue,~C.;\ \ Zheng,~X.;\ \ Chen,~K.;\ \ Tian,~Y.;\ \ Hu,~G. \textit{The Journal
  of Physical Chemistry Letters} \textbf{2016,} \textsl{7,} 514--519.

\bibitem{Ranjit_JPCB_2015}
Pal,~T.;\ \ Biswas,~R. \textit{The Journal of Physical Chemistry B}
  \textbf{2015,} \textsl{119,} 15683-15695.

\bibitem{Ranjit_JCP_2016}
Indra,~S.;\ \ Guchhait,~B.;\ \ Biswas,~R. \textit{The Journal of Chemical
  Physics} \textbf{2016,} \textsl{144,} 124506.

\bibitem{Sarika_PNAS_08}
Bhattacharyya,~S.~M.;\ \ Bagchi,~B.;\ \ Wolynes,~P.~G. \textit{Proceedings of
  the National Academy of Sciences} \textbf{2008,} \textsl{105,} 16077--16082.

\bibitem{Chen_JPCB_2010}
Mallamace,~F.;\ \ Branca,~C.;\ \ Corsaro,~C.;\ \ Leone,~N.;\ \ Spooren,~J.;\ \
  Stanley,~H.~E.;\ \ Chen,~S.-H. \textit{The Journal of Physical Chemistry B}
  \textbf{2010,} \textsl{114,} 1870-1878.

\bibitem{Gallo_JCP_2000}
Gallo,~P.;\ \ Rovere,~M.;\ \ Spohr,~E. \textit{The Journal of Chemical Physics}
  \textbf{2000,} \textsl{113,} 11324--11335.

\bibitem{Gallo_PRL_2000}
Gallo,~P.;\ \ Rovere,~M.;\ \ Spohr,~E. \textit{Physical Review Letter}
  \textbf{2000,} \textsl{85,} 4317--4320.

\bibitem{Funel_JML_1998}
Bellissent-Funel,~M.-C. \textit{Journal of Molecular Liquids} \textbf{1998,}
  \textsl{78,} 19--28.

\bibitem{Bilge_JACS_2011}
Mostafa,~Y.;\ \ J.-M.,~P.~R.;\ \ Bilge,~Y. \textit{Journal of the American
  Chemical Society} \textbf{2011,} \textsl{133,} 2499--2510.

\bibitem{Chen_PRE_1996}
Sciortino,~F.;\ \ Gallo,~P.;\ \ Tartaglia,~P.;\ \ Chen,~S.~H. \textit{Physical
  Review E} \textbf{1996,} \textsl{54,} 6331--6343.

\bibitem{SH_Chen_PRL_1996}
Gallo,~P.;\ \ Sciortino,~F.;\ \ Tartaglia,~P.;\ \ Chen,~S.-H. \textit{Physical
  Review Letter} \textbf{1996,} \textsl{76,} 2730--2733.

\bibitem{Eugene_PRL_2006}
Mazza,~M.~G.;\ \ Giovambattista,~N.;\ \ Starr,~F.~W.;\ \ Stanley,~H.~E.
  \textit{Physical Review Letter} \textbf{2006,} \textsl{96,} 057803.

\bibitem{Biman_JPCB_09}
Jana,~B.;\ \ Bagchi,~B. \textit{The Journal of Physical Chemistry B}
  \textbf{2009,} \textsl{113,} 2221--2224.

\bibitem{JLF_JCP_05}
Abascal,~J. L.~F.;\ \ Vega,~C. \textit{The Journal of Chemical Physics}
  \textbf{2005,} \textsl{123,} 234505-1--12.

\bibitem{Ananya_JCP_2010}
Debnath,~A.;\ \ Mukherjee,~B.;\ \ Ayappa,~K.~G.;\ \ Maiti,~P.~K.;\ \ Lin,~S.-T.
  \textit{The Journal of Chemical Physics} \textbf{2010,} \textsl{133,}
  174704-1--14.

\bibitem{Stillinger_JPCM_2005}
Shell,~M.~S.;\ \ Debenedetti,~P.~G.;\ \ Stillinger,~F.~H. \textit{Journal of
  Physics: Condensed Matter} \textbf{2005,} \textsl{17,} S4035.

\bibitem{Hansen_book}
Hansen,~J.-P.;\ \ McDonald,~I.~R. \textit{Theory of Simple Liquids (Fourth
  Edition);} 2013.

\bibitem{MAJ_PRE_2007}
Vorselaars,~B.;\ \ Lyulin,~A.~V.;\ \ Karatasos,~K.;\ \ Michels,~M. A.~J.
  \textit{Physical Review E} \textbf{2007,} \textsl{75,} 011504.

\bibitem{Walter_Kob_PRL_1997}
Kob,~W.;\ \ Donati,~C.;\ \ Plimpton,~S.~J.;\ \ Poole,~P.~H.;\ \ Glotzer,~S.~C.
  \textit{Physical Review Letter} \textbf{1997,} \textsl{79,} 2827--2830.

\bibitem{Schmidt_JCP_2010}
Hopkins,~P.;\ \ Fortini,~A.;\ \ Archer,~A.~J.;\ \ Schmidt,~M. \textit{The
  Journal of Chemical Physics} \textbf{2010,} \textsl{133,} 224505.

\bibitem{Pinaki_PRL_07}
Chaudhuri,~P.;\ \ Berthier,~L.;\ \ Kob,~W. \textit{Physical Review Letter}
  \textbf{2007,} \textsl{99,} 060604.

\bibitem{Siladitya_JCP_14}
Sengupta,~S.;\ \ Karmakar,~S. \textit{The Journal of Chemical Physics}
  \textbf{2014,} \textsl{140,} 224505.

\bibitem{Srikanth_PRL_17}
Parmar,~A. D.~S.;\ \ Sengupta,~S.;\ \ Sastry,~S. \textit{Physical Review
  Letter} \textbf{2017,} \textsl{119,} 056001.

\bibitem{Gallo_JPCL_2011}
Magno,~A.;\ \ Gallo,~P. \textit{The Journal of Physical Chemistry Letters}
  \textbf{2011,} \textsl{2,} 977--982.

\bibitem{Gallo_JCP_2016}
Camisasca,~G.;\ \ Marzio,~M.~D.;\ \ Corradini,~D.;\ \ Gallo,~P. \textit{The
  Journal of Chemical Physics} \textbf{2016,} \textsl{145,} 044503.

\bibitem{Smarajit_JStatMech_16}
Bhowmik,~B.~P.;\ \ Das,~R.;\ \ Karmakar,~S. \textit{Journal of Statistical
  Mechanics: Theory and Experiment} \textbf{2016,} \textsl{2016,} 074003.

\bibitem{Das_JCP_13}
Das,~J.;\ \ Flenner,~E.;\ \ Kosztin,~I. \textit{The Journal of Chemical
  Physics} \textbf{2013,} \textsl{139,} 065102.

\bibitem{Sharon_PRL_1997}
Kob,~W.;\ \ Donati,~C.;\ \ Plimpton,~S.~J.;\ \ Poole,~P.~H.;\ \ Glotzer,~S.~C.
  \textit{Physical Review Letter} \textbf{1997,} \textsl{79,} 2827--2830.

\bibitem{Ritchie_Biophy_2005}
Ritchie,~K.;\ \ Shan,~X.-Y.;\ \ Kondo,~J.;\ \ Iwasawa,~K.;\ \ Fujiwara,~T.;\ \
  Kusumi,~A. \textit{Biophysical Journal} \textbf{2005,} \textsl{88,}
  2266--2277.
 
 \bibitem{Berendsen_JCP_84}
Berendsen,~H. J.~C.;\ \ Postma,~J. P.~M.;\ \ van Gunsteren,~W.~F.;\ \
  DiNola,~A.;\ \ Haak,~J.~R. \textit{The Journal of Chemical Physics}
  \textbf{1984,} \textsl{81,} 3684--3690.

 \bibitem{Ulrich_JCP_95}
Essmann,~U.;\ \ Perera,~L.;\ \ Berkowitz,~M.~L.;\ \ Darden,~T.;\ \ Lee,~H.;\ \
  Pedersen,~L.~G. \textit{The Journal of Chemical Physics} \textbf{1995,}
  \textsl{103,} 8577--8593. 
  
  \bibitem{Allen}
Allen~M.P.,~T.~D.  \textbf{1987,} .

\bibitem{Darden_JCP_93}
Darden,~T.;\ \ York,~D.;\ \ Pedersen,~L. \textit{The Journal of Chemical
  Physics} \textbf{1993,} \textsl{98,} 10089--10092.

  \bibitem{RENARDUS_PC_93}
Bekker,~H.;\ \ Berendsen,~H.;\ \ Dijkstra,~E.;\ \ Achterop,~S.;\ \
  Vondrumen,~R.;\ \ van~der Spoel,~D.;\ \ Sijbers,~A.;\ \ Keegstra,~H.;\ \
  Renardus,~M.  \textbf{1993,}  252--256.
  
\bibitem{Hess_MMA_01}
Lindahl,~E.;\ \ Hess,~B.;\ \ van~der Spoel,~D. \textit{Molecular modelling
  annual} \textbf{2001,} \textsl{7,} 306--317.

  \bibitem{Berger_Biophys_97}
Berger,~O.;\ \ Edholm,~O.;\ \ Jähnig,~F. \textit{Biophysical Journal}
  \textbf{1997,} \textsl{72,} 2002--2013.

\bibitem{Cordomí_JCTC_12}
Cordomí,~A.;\ \ Caltabiano,~G.;\ \ Pardo,~L. \textit{Journal of Chemical
  Theory and Computation} \textbf{2012,} \textsl{8,} 948--958.
  
\bibitem{Erik_JCTC_08}
Berk Hess, Carsten Kutzner, David van~der Spoel, and Erik Lindahl.
\newblock Gromacs 4:  algorithms for highly efficient, load-balanced, and
  scalable molecular simulation.
\newblock {\em Journal of Chemical Theory and Computation}, 4(3):435--447,
  2008.
\newblock PMID: 26620784.

\bibitem{JCH_JCC_05}
David Van Der~Spoel, Erik Lindahl, Berk Hess, Gerrit Groenhof, Alan~E. Mark,
  and Herman J.~C. Berendsen.
\newblock Gromacs: Fast, flexible, and free.
\newblock {\em Journal of Computational Chemistry}, 26(16):1701--1718, 2005.

\bibitem{Hess_2013}
B.~Hess D.~van~der Spoel, E.~Lindahl and the GROMACS~development team.
\newblock Gromacs user manual version 4.6.5.
\newblock 2013.

\bibitem{Drunen_CPC_95}
H.J.C. Berendsen, D.~van~der Spoel, and R.~van Drunen.
\newblock Gromacs - a message-passing parallel molecular-dynamics
  implementation.
\newblock {\em Computer Physics Communications}, 91(1-3):43--56, 9 1995.
\newblock Journal EP F322 OMPUT PHYS COMMUN.
  


  
  
\end{thebibliography}
% \end{thebibliography}
\end{document}